\documentclass[11pt,a4paper,english]{article}
\usepackage[round,sort]{natbib}
\usepackage[latin1]{inputenc}
\usepackage[T1]{fontenc} 
\usepackage[english]{babel}
\usepackage{amsmath,amsfonts,dsfont,amsthm,amssymb,amsbsy,graphicx,graphics,pdfpages, geometry}
\usepackage{subfigure}
\usepackage{booktabs}
\newcommand\T{\rule{0pt}{2.6ex}}       % Top strut
\newcommand\B{\rule[-1.2ex]{0pt}{0pt}} % Bottom strut

\theoremstyle{definition}

\newtheorem{scheme}{Scheme}

\title{\textbf{Combining low-dimensional ensemble postprocessing with reordering methods}}
\author{Roman Schefzik\vspace{0.2 cm}\\ \textit{Heidelberg Institute for Theoretical Studies}\\ \textit{Schloss-Wolfsbrunnenweg 35, 69118 Heidelberg, Germany}\vspace{0.2 cm}\\ \texttt{roman.schefzik@h-its.org}}
\date{}
\begin{document}

%% Necessary!
\maketitle

\begin{abstract}
State-of-the-art weather forecasts usually rely on ensemble prediction systems, accounting for the different sources of uncertainty. As ensembles are typically uncalibrated, they should get statistically postprocessed. Several multivariate ensemble postprocessing techniques, which additionally consider spatial, inter-variable and/or temporal dependencies, have been developed. These can be roughly divided into two groups. The first group comprises parametric, mainly low-dimensional approaches that are tailored to specific settings. The second group involves non-parametric reordering methods that impose a specific dependence template on univariately postprocesed forecasts and are suitable in any dimension. In this paper, these different strategies are combined, with the aim to exploit the benefits of both concepts. Specifically, a high-dimensional postprocessing problem is divided into multiple low-dimensional instances, each of which is postprocessed via a suitable multivariate parametric method. From each postprocessed low-dimensional distribution, a sample is drawn, which is then reordered according to the corresponding multidimensional rank structure of an appropriately chosen dependence template. In this context, different ranking concepts for multivariate settings are discussed. Finally, all reordered samples are aggregated to obtain the overall postprocessed ensemble. The new approach is applied to ensemble forecasts for temperature and wind speed at several locations from the European Centre for Medium-Range Weather Forecasts, using a recent bivariate ensemble model output statistics postprocessing technique and a reordering based on the raw ensemble forecasts similar to the ensemble copula coupling method. It shows good predictive skill and outperforms reference ensembles.
\end{abstract}


%%%%%%%%%%%%%%%%%%%%%%%%%%%%%%%%%%%%%%%%%%%%%%%%%%%%%%%%%%%%%%%%%%%%%
% MAIN BODY OF PAPER
%%%%%%%%%%%%%%%%%%%%%%%%%%%%%%%%%%%%%%%%%%%%%%%%%%%%%%%%%%%%%%%%%%%%%
%

%% In all cases, if there is only one entry of this type within
%% the higher level heading, use the star form: 
%%
% \section{Section title}
% \subsection*{subsection}
% text...
% \section{Section title}

%vs

% \section{Section title}
% \subsection{subsection one}
% text...
% \subsection{subsection two}
% \section{Section title}

%%%
% \section{First primary heading}

% \subsection{First secondary heading}

% \subsubsection{First tertiary heading}

% \paragraph{First quaternary heading}

\section{Introduction}\label{intro}

Nowadays, weather forecasts are typically derived from ensemble prediction systems, which comprise different runs of numerical weather prediction models varying in terms of the initial conditions and/or the parameterization of the atmosphere \citep{Palmer2002,GneitingRaftery2005,LeutbecherPalmer2008}. While ensembles thus account for the main sources of predictive uncertainty, they tend to exhibit biases and lack calibration \citep{HamillColucci1997}, in that the verifying observation falls too often outside the ensemble range \citep{Gneiting&2007}. Consequently, an ensemble forecast should get statistically postprocessed. Numerous ensemble postprocessing methods have been introduced during the last decade, with the ensemble model output statistics (EMOS) \citep[among others]{Gneiting&2005,ThorarinsdottirGneiting2010} approach and Bayesian model averaging (BMA) \citep[among others]{Raftery&2005,Baran2014} being two key examples. However, most of these techniques, including standard EMOS and BMA, do not account for potentially crucial dependence structures in space or time, or among different weather quantities.
\\
To address this, a lot of work has been done on multivariate postprocessing approaches modeling correlation patterns adequately, in particular in recent years. Among others, variants of EMOS and BMA, respectively, which are able to deal with spatial \citep{Berrocal&2007,Berrocal&2008,Feldmann&2014} or inter-variable \citep{Schuhen&2012,Moeller&2013,Sloughter&2013,BaranMoeller2015,BaranMoeller2014} dependencies have been proposed. Moreover, temporal dependencies of consecutive lead times in postprocessed predictive distributions can be treated via the approaches of \citet{Pinson&2009} or \citet{SchoelzelHense2011}. The aforementioned multivariate postprocessing techniques are parametric and perform well in low-dimensional settings, or when the involved correlation matrix can be assumed to be highly structured.
\\
However, they mostly address either spatial or inter-variable or temporal dependencies only and are not suitable for high-dimensional settings without any specific structure. In such cases, multivariate ensemble postprocessing can be appropriately done using non-parametric reordering methods \citep{Wilks2014,Schefzik2015c}, which are conceptionally simpler than the parametric approaches. Essentially, these methods rearrange for each margin a sample drawn from the respective predictive distribution obtained by univariate ensemble postprocessing (via EMOS or BMA, for instance) according to the rank dependence structure of a suitable dependence template \citep{Wilks2014,Schefzik2015c}. Aggregating these reordered samples yields the postprocessed ensemble respecting dependence patterns. Examples of reordering methods include the Schaake shuffle \citep{Clark&2004}, as well as specific implementations thereof \citep{Schefzik2015c}, and ensemble copula coupling (ECC) \citep{Schefzik&2013}. While the Schaake shuffle employs a database of historical observations as a dependence template, ECC relies on the information in the raw ensemble forecast instead. From a mathematical point of view, reordering methods can be interpreted in the general frame of empirical copulas, as discussed by \citet{Schefzik2015b}.
\\
In this paper, we develop a new approach which combines parametric Low-Dimensional Postprocessing (LDP) methods tailored to specific settings on the one hand with the non-parametric Reordering notion on the other hand, with the goal of exploiting the benefits of both concepts. It will be referred to as the LDP-Reordering approach in what follows. In the context of this new technique, the question of how to define a ranking concept in multivariate settings comes into play. The course of action in the paper is two-part. First, the LDP-Reordering method is presented in a very general frame, being broadly applicable. Second, we consider an example, in which the bivariate EMOS model of \citet{BaranMoeller2015} is used for the joint postprocessing of 10 meter wind speed and surface temperature forecasts at distinct stations individually. The respective postprocessed bivariate samples are then reordered and aggregated via an ECC-like approach using an appropriate reordering notion for multivariate settings. We refer to this specific implementation, which will be tested and evaluated in a case study exemplarily for an LDP-Reordering method, as the Bivariate EMOS-ECC approach. 
\\
The remainder of this paper is organized as follows. In view of the aforementioned illustrative example and the later case study, we review in Section \ref{prelim} relevant univariate and bivariate EMOS approaches, as well as the concept of reordering methods, with an emphasis on ECC. Moreover, notions of ranking in multivariate settings are discussed. In Section \ref{ldp.ecc}, we first introduce the general LDP-Reordering approach and then consider Bivariate EMOS-ECC as an example. Section \ref{case.study} deals with the evaluation of the Bivariate EMOS-ECC approach in a case study using real forecast data from the European Centre for Medium-Range Weather Forecasts. Finally, the paper closes with a discussion in Section \ref{discussion}.

\section{Preliminaries}\label{prelim}

In this preparatory section, we review existing postprocessing methods that are relevant in our case study. In addition, different ranking concepts in multivariate settings are elucidated.

\subsection{Reference ensemble postprocessing methods}\label{ref.methods}

\subsubsection{Ensemble model output statistics (EMOS)}\label{emos}

The standard univariate ensemble model output statistics (EMOS) approach, also known as non-homogeneous regression, fits a single probability distribution for a fixed weather quantity at a fixed location and for a fixed look-ahead time, using summary statistics from a raw ensemble forecast $x_1,\ldots,x_M$. The choice of the specific type of the predictive distribution depends on the considered weather variable. For instance, EMOS variants are available for temperature and pressure \citep{Gneiting&2005}, wind speed \citep{ThorarinsdottirGneiting2010,LerchThorarinsdottir2013,BaranLerch2014} or precipitation \citep{Scheuerer2013}.    
\\
In the case of temperature $y$, EMOS employs a normal predictive distribution
\begin{equation}\label{emos.gaussian}
{\cal{N}}(a+b_{1}x_1+\ldots+b_{M}x_M,c+ds^2)
\end{equation}
having mean $a+b_1x_1+\ldots+b_Mx_M$ and variance $c+ds^2$, where $a,b_1,\ldots,b_M,c$ and $d$ are parameters that need to be estimated, and $s^2:=(1/M)\sum_{m=1}^{M}(x_m-\bar{x})^2$ denotes the empirical ensemble variance, with the empirical ensemble mean $\bar{x}:=(1/M)\sum_{m=1}^{M}x_m$ \citep{Gneiting&2005}. While $a$ is a bias correction term, the regression coefficients $b_1,\ldots,b_M$ reflect the performance of the ensemble members over the training period relative to the other members, as well as the correlations between the ensemble members \citep{Gneiting&2005}. If the ensemble members are statistically indistinguishable and can thus be considered exchangeable, the regression coefficients are assumed to be equal, such that $b_1=\cdots=b_M$.
\\
For wind speed $y$, which is a nonnegative quantity, we follow \citet{ThorarinsdottirGneiting2010} and use a truncated normal distribution with a cut-off at zero, that is, 
\begin{equation}\label{emos.truncnormal}
{\cal{N}}^{0}(a+b_{1}x_1+\ldots+b_{M}x_M,c+ds^2),
\end{equation}
with parameters as in \eqref{emos.gaussian}.\\
Both in \eqref{emos.gaussian} and \eqref{emos.truncnormal}, the parameters $a,b_1,\ldots,b_M,c$ and $d$ are estimated by optimizing a proper scoring rule \citep{GneitingRaftery2007} over a training data set consisting of past forecasts and observations. According to \citet{Gneiting&2005}, common choices for the proper scoring rule to be optimized are the logarithmic (or ignorance) score \citep{Good1952,RoulstonSmith2002}, following the classical maximum likelihood notion, and the continuous ranked probability score \citep{Hersbach2000,GneitingRaftery2007}. For both scores, closed-form expressions are available in the above two cases.
\\
The EMOS approach for temperature is implemented in the R package {\texttt{ensembleMOS}} \citep{R2011,Yuen&2013}.
\\
To get an ensemble representation of a full EMOS predictive cumulative distribution function (CDF) $F$, we draw a sample $\tilde{x}_1,\ldots,\tilde{x}_N$ thereof. This can be for instance done by taking the equally spaced quantiles
\begin{equation}\tag{Q}\label{emos.q}
\tilde{x}_{1}:=F^{-1}\left(\frac{1}{N+1}\right),\ldots,\tilde{x}_{N}:=F^{-1}\left(\frac{N}{N+1}\right).
\end{equation}
Alternatively, we can just sample randomly from $F$, that is,
\begin{equation}\tag{R}\label{emos.r}
\tilde{x}_{1}:=F^{-1}(u_1),\ldots,\tilde{x}_{N}:=F^{-1}(u_N).
\end{equation}
where $u_1,\ldots,u_N$ are independent standard uniform random variates. 
\\
The corresponding ensembles are referred to as EMOS-Q and EMOS-R ensembles, respectively, in this paper, following a similar notation as in \citet{Schefzik&2013}.

\subsubsection{Bivariate EMOS for joint postprocessing of wind speed and temperature forecasts}\label{lowdim.postpr}

For the joint postprocessing of non-negative wind speed and real-valued temperature forecasts at a fixed location and for a fixed prediction horizon, \citet{BaranMoeller2015} propose to use a bivariate normal distribution $\mathcal{N}_2^0({\boldsymbol{\mu}},\boldsymbol{\Sigma})$ with the first coordinate truncated from below at zero, where $\boldsymbol{\mu}$ is a location vector and $\boldsymbol{\Sigma}$ a scale matrix. Precisely, if $\mathbf{x}_1:=(x_1^W,x_1^T),\ldots,\mathbf{x}_M:=(x_M^W,x_M^T)$ denotes the bivariate $M$-member raw ensemble forecast for wind speed (W) and temperature (T), the  corresponding EMOS predictive distribution is given by 
\begin{equation}\label{bivar.emos}
\mathcal{N}_2^0(\mathbf{A}+\mathbf{B}_1\mathbf{x}_1+\ldots+\mathbf{B}_M\mathbf{x}_M,\mathbf{C}+\mathbf{DSD}^{\prime}),
\end{equation}
where 
\begin{equation*}
\mathbf{S}:=\frac{1}{M-1}\sum\limits_{m=1}^{M}(\mathbf{x}_m-\mathbf{\bar{x}})(\mathbf{x}_m-\mathbf{\bar{x}})^{\prime},
\end{equation*}
with $\mathbf{\bar{x}}$ being the empirical ensemble mean vector and $\boldsymbol{\Upsilon}^{\prime}$ denoting the transpose of the matrix or vector $\boldsymbol{\Upsilon}$. In \eqref{bivar.emos}, $\mathbf{A}$ is a bivariate parameter vector, and $\mathbf{B}_1,\ldots,\mathbf{B}_M,\mathbf{C}$ and $\mathbf{D}$ are two-by-two real parameter matrices, with $\mathbf{C}$ being symmetric and non-negative definite. \citet{BaranMoeller2015} estimate the parameters by optimizing with respect to the mean logarithmic score, using past ensemble forecasts and observations as training data. For details, see \citet{BaranMoeller2015}. 
\\
To sample from a bivariate normal distribution with the first coordinate truncated from below at zero, one can use a rejection sampling method, which draws from an untruncated bivariate normal distribution and accepts only those samples that lie inside the support region. Alternatively, a Gibbs sampler can be employed. Both sampling options are implemented in the R package {\texttt{tmvtnorm}} \citep{WilhelmManjunath2010,WilhelmManjunath2014}.

\subsubsection{Reordering methods, in particular ensemble copula coupling (ECC)}\label{ecc}

The bivariate EMOS method from \ref{lowdim.postpr} joins the steadily growing list of purely parametric ensemble postprocessing techniques tailored to specific low-dimensional settings \citep[among others]{Schuhen&2012,Pinson2012,BaranMoeller2014,Moeller&2013}. However, parametric approaches are often inadequate in high-dimensional settings and fail to address spatial, inter-variable and temporal dependence structures simultaneously. In contrast, non-parametric methods based on reordering notions provide suitable and appealing tools to model such dependencies together, also in higher dimensions.
\\
For the following general description of reordering methods, we stick to Scheme 1 in \citet{Schefzik2015c}. Let $\ell$ be a fixed margin, summarizing a fixed weather quantity, a fixed location and a fixed prediction horizon, and let $L$ be the total number of margins. The goal is to get a postprocessed ensemble forecast $\mathbf{\hat{x}}:=\{(\hat{x}_{1}^{1},\ldots,\hat{x}_{N}^{1}),\ldots,(\hat{x}_{1}^{L},\ldots,\hat{x}_{N}^{L})\}$ out of an $M$-member raw ensemble forecast $\mathbf{x}:=\{(x_{1}^{1},\ldots,x_{M}^{1}),\ldots,(x_{1}^{L},\ldots,x_{M}^{L})\}$. In a first step, we choose a suitable data set ${\mathbf{z}}:=\{(z_{1}^{1},\ldots,z_{N}^{1}),\ldots,(z_{1}^{L},\ldots,z_{N}^{L})\}$ which serves as a dependence template \citep{Wilks2014} and adequately models correlation structures among the $L$ margins. For each margin $\ell \in \{1,\ldots,L\}$, we derive the univariate order statistics 
\begin{equation}\label{univ.order}
z_{(1)}^{\ell} \leq \ldots \leq z_{(N)}^{\ell},
\end{equation}
inducing the permutation $\pi_{\ell}(n):=\operatorname{rank}(z_n^{\ell})$ for $n \in \{1,\ldots,N\}$, with ties resolved at random. Then, univariate postprocessing of the raw ensemble forecast $x_1^\ell,\ldots,x_M^\ell$ is performed for each margin $\ell$, for instance via EMOS, yielding a postprocessed predictive CDF $F_\ell$. From each marginal CDF $F_\ell$, we draw a sample $\tilde{x}_{1}^{\ell},\ldots,\tilde{x}_{N}^{\ell}$, for example by using one of the schemes (Q) or (R) from \ref{emos}. Finally, we arrange these samples with respect to the rank dependence structure of ${\mathbf{z}}$. Hence, using the permutation $\pi_\ell$, the final postprocessed ensemble $\hat{x}_{1}^{\ell},\ldots,\hat{x}_{N}^{\ell}$ for each margin $\ell$ is given by
\begin{equation*}
\hat{x}_{1}^{\ell}:=\tilde{x}_{(\pi_\ell(1))}^{\ell},\ldots,\hat{x}_{N}^{\ell}:=\tilde{x}_{(\pi_\ell(N))}^{\ell}.
\end{equation*}
Aggregating the $L$ postprocessed margins yields the desired overall postprocessed ensemble $\mathbf{\hat{x}}$. From a mathematical point of view, reordering methods can be interpreted in the frame of discrete or empirical copulas \citep{Wilks2014,Schefzik2015b,Schefzik2015c}.
\\
There are several options for the specific choice of the dependence template $\mathbf{z}$.
\\
In the Schaake shuffle \citep{Clark&2004}, $\mathbf{z}$ is based on past observations from a historical database. A variant of the Schaake shuffle particularly employs historical observations from dates on which the corresponding ensemble forecast resembled the current one \citep{Schefzik2015c}.
\\  
In the ECC approach \citep{Schefzik&2013}, $\mathbf{z}$ is given by the raw ensemble forecast, that is, $\mathbf{z}=\mathbf{x}$. Hence, the size of the final postprocessed ensemble equals that of the raw ensemble, that is, $N=M$. ECC implicitly assumes that the ensemble is able to represent actual spatial, inter-variable and temporal dependence structures appropriately. This is surely not valid each and every day, but may largely be expected. Furthermore, ECC only applies to ensembles whose members are statistically indistinguishable and thus can be considered exchangeable.
\\
In the special case of using univariate postprocessing via EMOS, together with the discretization schemes (Q) and (R), respectively, in the ECC approach, we refer to the corresponding ensembles as EMOS-ECC-Q and EMOS-ECC-R.

\subsection{Ranking in multivariate settings}\label{mult.ordering}

Ranking $N$ one-dimensional data points $z_1,\ldots,z_N \in \mathbb{R}$ can be conveniently done by using the usual univariate ordering $z_{(1)} \leq \cdots \leq z_{(N)}$ as in \eqref{univ.order}, defining the permutation $\pi(n):=\operatorname{rank}(z_n)$ for $n \in \{1,\ldots,N\}$, with ties resolved at random. The situation in the multivariate case is however more involved.
\\
To rank $N$ data points $\mathbf{z}_1:=(z_1^1,\ldots,z_1^L),\ldots,\mathbf{z}_N:=(z_N^1,\ldots,z_N^L) \in \mathbb{R}^L$ with $L$-dimensional output each, we proceed in the following steps:
\begin{scheme}\label{scheme.multor}(Ranking in multivariate settings)
\begin{enumerate}
\item{Each data point $\mathbf{z}_n$ with index $n \in \{1,\ldots,N\}$ is assigned a suitable multivariate characteristic $R_n \in \mathbb{R}$. There are several options to choose $R_n$, as will be discussed below.}
\item{We rank the values $R_1,\ldots,R_N$ by using the common univariate ordering $R_{(1)} \leq \cdots \leq R_{(N)}$, inducing the permutation $\tau_n:=\operatorname{rank}(R_n)$ for $n \in \{1,\ldots,N\}$, with ties resolved at random.}
\item{Each data point $\mathbf{z}_n$ for $n \in \{1,\ldots,N\}$ is finally assigned the corresponding rank $\tau_n$ derived in step 2.}
\end{enumerate}
\end{scheme}
\noindent As mentioned, there are several possibilities to define the multivariate characteristic $R_n$ the ranking is based on. For instance, we can use the multivariate pre-rank construction following \citet{Gneiting&2008} and set
\begin{equation}\label{multrank}
R_{n}^{\operatorname{mult}}:=\sum\limits_{\nu=1}^{N} \mathds{1}_{\{\mathbf{z}_{\nu} \preceq \mathbf{z}_{n}\}},
\end{equation}
where $\mathbf{z}_{\nu} \preceq \mathbf{z}_{n}$ if and only if $z_{\nu}^\ell \leq z_{n}^\ell$ for all $\ell \in \{1,\ldots,L\}$, and $\mathds{1}_A$ denotes the indicator function of the event $A$ which is 1 if $A$ occurs and 0 otherwise. This concept works well in low-dimensional settings, as in the bivariate scenario in our case study, but may face problems when dealing with high dimensions \citep{Thorarinsdottir&2013}. Another option to define a multivariate characteristic may employ the average pre-rank as introduced by \citet{Thorarinsdottir&2013}, which is especially suitable in high-dimensional settings. The average pre-rank 
\begin{equation}\label{avgrank}
R_n^{\operatorname{avg}}:=\frac{1}{L}\sum\limits_{\ell=1}^{L}\operatorname{rank}_S(z_n^\ell)
\end{equation}
is just the average over the univariate ranks, with $\operatorname{rank}_S(z_n^\ell)=\sum_{\nu=1}^{N}\mathds{1}_{\{z_\nu^\ell \leq z_n^\ell\}}$ denoting the rank of the $\ell$-th coordinate of $\mathbf{z}_n$ in $S:=\{\mathbf{z}_1,\ldots,\mathbf{z}_N\}$ for $n \in \{1,\ldots,N\}$ \citep{Thorarinsdottir&2013}. Further, alternative possibilities to define a multivariate characteristic may be based on minimum spanning trees \citep{Wilks2004,SmithHansen2004,Gombos&2007} or the band depth pre-rank \citep{Thorarinsdottir&2013}.
\\
While the above suggestions are rather general, we propose in the following a multivariate characteristic that is especially tailored to the setting of non-negative wind speed and real-valued temperature data in our case study.  For wind speed (W) data points  $z_1^W,\ldots,z_N^W \in \mathbb{R}_0^{+}$ and temperature (T) data points $z_1^T,\ldots,z_N^T \in \mathbb{R}$, which should ideally all be standardized, we define what can be considered a signed Euclidean norm as follows:
\begin{equation}\label{norms.signbased}
R_{n}^{\operatorname{Eucl}}:=\operatorname{sgn}(z_n^T) \cdot \sqrt{(z_n^W)^2+(z_n^T)^2}
\end{equation}
for $n \in \{1,\ldots,N\}$, where
\begin{equation*}
\operatorname{sgn}(z_n^T)=\begin{cases}1& \operatorname{if\,\,\,} z_n^T \geq 0 \\ -1 & \operatorname{if\,\,\,} z_n^T <0. \end{cases}
\end{equation*}
Again, this concept is specifically tailored to the setting of our case study and thus should not be used without further reflection as a general option in other settings.
\begin{figure*}[t]
\centering
\includegraphics[scale=0.48]{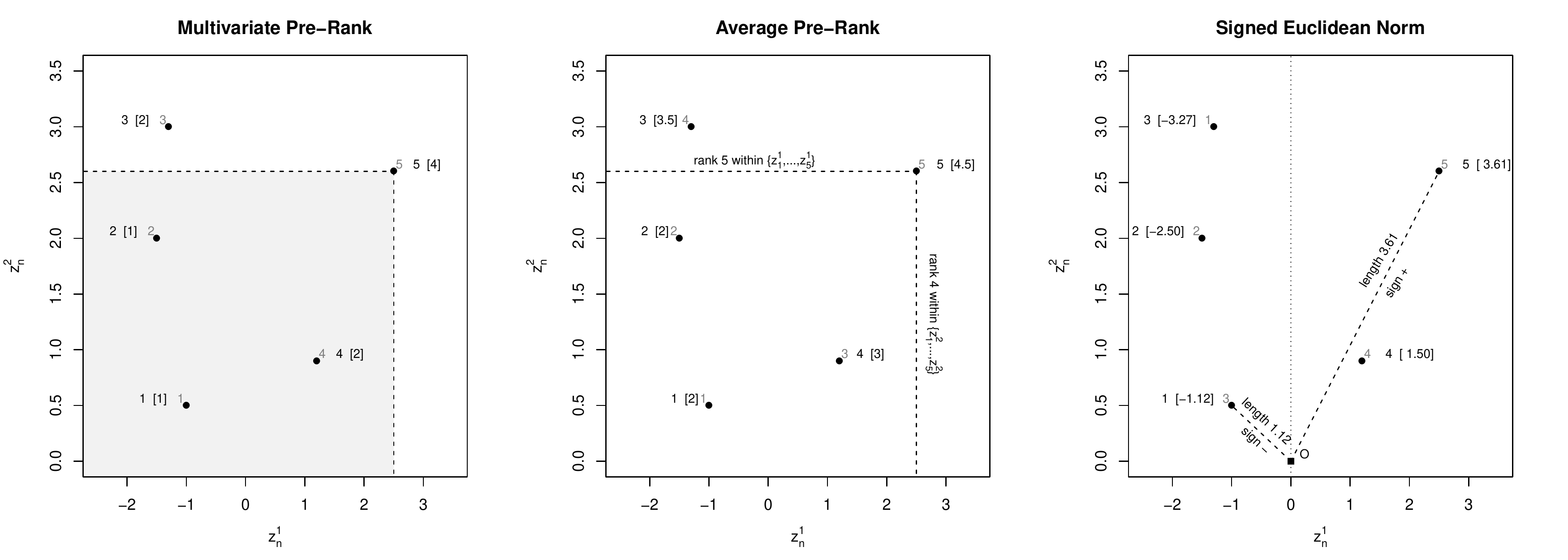}
\caption{Illustration of multivariate ranking based on three different multivariate characteristics as discussed in Subsection \ref{mult.ordering}, namely the multivariate pre-rank as in \eqref{multrank}, the average pre-rank as in \eqref{avgrank} and the signed Euclidean norm as in \eqref{norms.signbased}. Each data point indicated by a dot is labeled in black with its index and its corresponding multivariate characteristic in brackets. The number in gray indicates the final rank.}
  \label{fig:illus.multord}
\end{figure*}
\\
\noindent An illustration of the multivariate ranking approaches based on the multivariate characteristics \eqref{multrank}, \eqref{avgrank} and \eqref{norms.signbased}, respectively, is given in Figure \ref{fig:illus.multord}. In each plot, we show $N=5$ data points ${\mathbf{z}}_1:=(z_1^1,z_1^2),\ldots,{\mathbf{z}}_5:=(z_5^1,z_5^2)$ of dimension $L=2$ each. With regard to the case study, we can think of $\mathbf{z}_1,\ldots,\mathbf{z}_5$ as ensemble member forecasts, where the $z_n^1$'s can be considered temperature (T) forecasts, and the $z_n^2$'s wind speed (W) forecasts for $n \in\{1,\ldots,5\}$, thus $z_n^1=z_n^T$ and $z_n^2=z_n^W$. Each data point $\mathbf{z}_n$ indicated by a dot is in each case labeled in black with its corresponding index $n$, ranging from 1 to 5, and its associated multivariate (bivariate) characteristic $R_n$ in brackets. The final rank $\tau_n$ of each data point $\mathbf{z}_n$ according to Scheme \ref{scheme.multor}, ranging from 1 to 5 and obtained by ranking the multivariate characteristics according to common univariate ordering, is labeled in gray.\\
The plot at left in Figure \ref{fig:illus.multord} shows the ranking following the multivariate pre-rank idea in \eqref{multrank}. For instance, the data point $\mathbf{z}_5$ with the index 5 has bivariate pre-rank $R_5^{\operatorname{mult}}=4$ according to \eqref{multrank}, as 4 of the 5 data points lie to its lower left, including $\mathbf{z}_5$ itself. As this is the highest among all bivariate pre-ranks, $\mathbf{z}_5$ is assigned the final rank 5.\\
In the middle plot, the ranking based on the use of average pre-ranks according to \eqref{avgrank} is illustrated. Exemplarily, the data point $\mathbf{z}_5$ has an average pre-rank of 4.5, as $z_5^1$ has rank 5 within the set $\{z_1^1,\ldots,z_5^1\}$ and $z_5^2$ has rank 4 within the set $\{z_1^2,\ldots,z_5^2\}$, such that $R_5^{\operatorname{avg}}=(5+4)/2=4.5$. Again, this is the highest value among all pre-ranks, such that $\mathbf{z}_5$ is assigned rank 5.\\
Finally, the plot at right exemplifies the ranking procedure using the signed Euclidean norm in \eqref{norms.signbased}, in which the distance from the data point $\mathbf{z}_n$ to the origin $O:=(0,0)$ is combined with a positive or negative sign, depending on whether $z_n^1 \geq 0$ or $z_n^1 < 0$, respectively. For example, the data point $\mathbf{z}_5$ has a signed Euclidean norm of 3.61, as the distance from $\mathbf{z}_5$ to $O$ is 3.61 and $z_5^1 \geq 0$. In contrast, $\mathbf{z}_1$ has a signed Euclidean norm of $-1.12$, since the distance from $\mathbf{z}_1$ to $O$ is $1.12$ and $z_1^1 < 0$. As $\mathbf{z}_1$ and $\mathbf{z}_5$ have the third highest and the highest signed Euclidean norms, respectively, they are assigned the final ranks 3 and 5, respectively.\\
Generally, for the signed Euclidean norm in \eqref{norms.signbased}, it is likely to obtain comparably few ties, if any, which reduces the need of a possible randomization in the ranking. In contrast, the chance of getting ties when using the multivariate pre-rank according to \eqref{multrank} appears to be higher, which also holds maybe to some lesser extent for the average pre-ranks in \eqref{avgrank}.

\section{Combining Low-Dimensional Postprocessing (LDP) and Reordering methods}\label{ldp.ecc}

In this section, we present a new ensemble postprocessing approach which combines parametric Low-Dimensional Postprocessing (LDP) methods and the non-parametric Reordering notion from \ref{ecc}. The method applies a multivariate reordering to low-dimensional building blocks and will be referred to as the LDP-Reordering approach in what follows. The goal is to exploit the benefits of both concepts with respect to an adequate modeling of dependence structures. We first decribe the LDP-Reordering method in a general frame being broadly applicable. In the second part of the section, we specifically discuss an example of the LDP-Reordering method dealing with the joint postprocessing of wind speed and temperature forecasts at different locations simultaneously. In this setting, which will also be used in the case study in Section \ref{case.study}, we combine the bivariate EMOS approach of \citet{BaranMoeller2015} from \ref{lowdim.postpr} with the ECC reordering notion of \citet{Schefzik&2013} from \ref{ecc} and refer to this procedure as the Bivariate EMOS-ECC method.

\subsection{The LDP-Reordering method}\label{ldp.empcop}

We first introduce the general LDP-Reordering method. Let  $i \in \{1,\ldots,I\}$ be a weather variable, $j \in \{1,\ldots,J\}$ a location and $k \in \{1,\ldots,K\}$ a look-ahead time, summarized in the multi-index $\ell:=(i,j,k)$, and let $L:=I \times J \times K$. Moreover, let 
\begin{equation*}
{\mathbf{x}}:=(\mathbf{x}_1,\ldots,\mathbf{x}_M)
\end{equation*}
be the overall $M$-member raw ensemble forecast which is to be postprocessed, where each ensemble member $m \in \{1,\ldots,M\}$ comprises an $L$-dimensional output 
\begin{equation*}
\mathbf{x}_m:=(x_{m}^{1},\ldots,x_{m}^{L}).
\end{equation*}
In cases comprising weather variables of different magnitudes and/or units, one should work with standardized forecast values in what follows. Suppose that $\mathbf{x}$ and $\mathbf{x}_m$ for each ensemble member $m$, respectively, is split up into $C$ sub-raw ensemble forecasts
\begin{equation*}
\mathbf{x}_m^c:=(x_m^{\ell^{c,1}},\ldots,x_m^{\ell^{c,L_c}})
\end{equation*}
where $c \in \{1,\ldots,C\}$, such that
\begin{equation*}
\mathbf{x}_m=(\mathbf{x}_m^1,\ldots,\mathbf{x}_m^{C}).
\end{equation*}
In this context, $\ell^{c,1},\ldots,\ell^{c,L_{c}}$ denote the $L_c$ multi-indices referring to those instances (weather variables, locations and/or look-ahead times) that shall be jointly postprocessed via a specific $L_c$-dimensional method in case $c \in \{1,\ldots,C\}$. In addition, $C$ is the number of cases in which a specific low-dimensional postprocessing method is intended to be applied, where the same approach may used multiple times. Moreover, $L_{c}$ is the dimension of the predictive CDF $F_{c}$ obtained by the corresponding postprocessing method in the case $c \in \{1,\ldots,C\}$. Thus, the overall dimension $L$ can be decomposed via
\begin{equation*}
L=\sum\limits_{c=1}^{C}L_{c}.
\end{equation*}
The specific choice of the involved LDP methods should be as plausible as possible and tailored to the setting one deals with. It might be reasonable to postprocess variables with an expected pronounced correlation jointly. For example, at individual stations, wind vectors could be postprocessed simultaneously via the bivariate EMOS approach of \citet{Schuhen&2012} or the method of \citet{Pinson2012},  while temperature and pressure could be treated jointly via the Gaussian copula approach of \citet{Moeller&2013}. On the other hand, purely spatial settings could be handled by Spatial BMA \citep{Berrocal&2007} or Spatial EMOS \citep{Feldmann&2014}, respectively. In our case study, we jointly postprocess wind speed and temperature forecasts via the bivariate EMOS method of \citet{BaranMoeller2015}. This specific example has been described in detail in \ref{lowdim.postpr}. 
\\
The general LDP-Reordering approach considered in this subsection proceeds in the following steps, with the goal to come up with a coherent, postprocessed ensemble forecast
\begin{equation*}
 \mathbf{\hat{x}}:=(\mathbf{\hat{x}}_{1},\ldots,\mathbf{\hat{x}}_{N}),
\end{equation*}
 with 
\begin{equation*}
\mathbf{\hat{x}}_{n}:=(\mathbf{\hat{x}}_{n}^{1},\ldots,\mathbf{\hat{x}}_{n}^{C})
\end{equation*}
for each postprocessed member $n \in \{1,\ldots,N\}$, where $N$ in general does not need to be equal to the raw ensemble size $M$.
\begin{enumerate}
\item[1.]{We choose a dependence template
\begin{equation*}
{\mathbf{z}}:=(\mathbf{z}_1,\ldots,\mathbf{z}_N)
\end{equation*}
based on $N$ data points, each of dimension $L$, where 
\begin{equation*}
\mathbf{z}_n:=(z_{n}^{1},\ldots,z_{n}^{L})
\end{equation*}
for an index  $n \in \{1,\ldots,N\}$. Similarly as above, we partition by setting
\begin{equation*}
\mathbf{z}_{n}^{c}:=(z_{n}^{\ell^{c,1}},\ldots,z_{n}^{\ell^{c,L_c}})
\end{equation*}
for case $c \in \{1,\ldots,C\}$, such that
\begin{equation*}
\mathbf{z}_{n}:=(\mathbf{z}_{n}^{1},\ldots,\mathbf{z}_{n}^{C}),
\end{equation*}
where $n \in \{1,\ldots,N\}$.
}
\end{enumerate}
The following steps 2 to 5 are applied to each case $c \in \{1,\ldots,C\}$ separately.
\begin{enumerate}
\item[2.]{
For each case $c$, we rank the $L_c$-dimensional points $\mathbf{z}_{1}^{c},\ldots,\mathbf{z}_{N}^{c}$ according to the scheme in Subsection \ref{mult.ordering}. That is, we choose first a suitable multivariate characteristic $R_n^c$ (with suggestions having been made in \ref{mult.ordering}), where the corresponding choice has to be consistent for all involved cases. Then, we derive the permutation $\tau^c$ given by $\tau^{c}(n):=\operatorname{rank}(R_{n}^{c})$ for $n \in \{1,\ldots,N\}$ using the usual univariate ordering, where ties are resolved at random.}
\item[3.]{We apply a reasonably chosen LDP method for the case $c$ to obtain an $L_{c}$-variate postprocessed predictive CDF $F_{c}$.
}
\item[4.]{From the predictive CDF $F_{c}$, we draw an $L_c$-dimensional sample $\mathbf{\tilde{x}}_{1}^{c},\ldots,\mathbf{\tilde{x}}_{N}^{c}$ of size $N$, where
\begin{equation*}
\mathbf{\tilde{x}}_{n}^{c}:=\left(\tilde{x}_{n}^{\ell^{c,1}},\ldots,\tilde{x}_{n}^{\ell^{c,L_{c}}}\right)
\end{equation*} 
for $n \in \{1,\ldots,N\}$.
}
\item[5.]{We reorder the sample $\mathbf{\tilde{x}}_{1}^{c},\ldots,\mathbf{\tilde{x}}_{N}^{c}$ from the previous step by using the permutation $\tau^{c}$ determined in step 2 to get the ensemble $\mathbf{\hat{x}}_{1}^{c},\ldots,\mathbf{\hat{x}}_{N}^{c}$, with
\begin{equation*}
\mathbf{\hat{x}}_{n}^{c}:=\left(\hat{x}_{n}^{\ell^{c,1}},\ldots,\hat{x}_{n}^{\ell^{c,L_{c}}}\right)
\end{equation*} 
for $n \in \{1,\ldots,N\}$. To this end, we derive by analogy with step 2 the multivariate characteristics $\tilde{R}_{1}^{c},\ldots,\tilde{R}_{N}^{c}$ associated to $\mathbf{\tilde{x}}_{1}^{c},\ldots,\mathbf{\tilde{x}}_{N}^{c}$. This induces the corresponding order statistics $\tilde{R}_{(1)_{\sim}}^{c} \leq \ldots \leq \tilde{R}_{(N)_{\sim}}^{c}$, with any ties resolved at random, using the notation $(\cdot)_{\sim}$ to emphasize that the ordering is based on the sample values. The corresponding permutation $\tilde{\tau}^{c}$ of $\{1,\ldots,N\}$ is then given by $\tilde{\tau}^{c}(n):=\operatorname{rank}(\tilde{R}_{n}^{c})$, and the reordered ensemble $\mathbf{\hat{x}}_{1}^{c},\ldots,\mathbf{\hat{x}}_{N}^{c}$ is finally obtained via
\begin{equation*}
\mathbf{\hat{x}}_{n}^{c}:=\mathbf{\tilde{x}}_{(\tilde{\tau}^{c})^{-1}(\tau^{c}(n))}^{c}
\end{equation*} 
for $n \in \{1,\ldots,N\}$. In other words using order statistic notation, $\mathbf{\hat{x}}_{1}^{c},\ldots,\mathbf{\hat{x}}_{N}^{c}$ is given by
\begin{equation*}
\mathbf{\hat{x}}_{n}^{c}:=\mathbf{\tilde{x}}_{(\tau^{c}(n))_{\sim}}^{c}
\end{equation*} 
for $n \in \{1,\ldots,N\}$. That is, the sample member with the index $(\tilde{\tau}^{c})^{-1}(\tau^{c}(n))$ from step 4 forms the member $n$ in the reordered ensemble. Therefore, the $L_c$-dimensional sample members $\mathbf{\tilde{x}}_{1}^{c},\ldots,\mathbf{\tilde{x}}_{N}^{c}$ from step 4 stay as they are with respect to their components, and only their order among themselves and hence the indices $n \in \{1,\ldots,N\}$ are changed, according to the multivariate ranking structure of the chosen dependence template.
}
\end{enumerate}
Having applied steps 2 to 5 to each case, we eventually aggregate all cases $c \in \{1,\ldots,C\}$ to get the final LDP-Reordering ensemble $\mathbf{\hat{x}}$.
\begin{enumerate}
\item[6.]{We first aggregate the postprocessed ensemble forecasts $\mathbf{\hat{x}}_{n}^{c}$ obtained in step 5 for all cases $c \in \{1,\ldots,C\}$ for each member $n \in \{1,\ldots,N\}$, which is done by setting
\begin{equation*}
\mathbf{\hat{x}}_{n}:=(\mathbf{\hat{x}}_{n}^{1},\ldots,\mathbf{\hat{x}}_{n}^{C}).
\end{equation*}
Finally, the LDP-Reordering ensemble $\mathbf{\hat{x}}$ is obtained via $\mathbf{\hat{x}}=(\mathbf{\hat{x}}_{1},\ldots,\mathbf{\hat{x}}_{N})$. 
}
\end{enumerate}
If a postprocessing method in case $c$ is univariate, such as standard EMOS as in \ref{emos}, the procedure reduces to a standard reordering approach.\\
For univariate methods, it may be most convenient and advisable to use equally spaced quantiles in the style of scheme (Q) in \ref{emos} in step 4. However, a generalization of the concept of equidistance to the multivariate case and its specific implementation are not straightforward. In our case study later, we thus employ random samples from the corresponding multivariate predictive distributions. The development of optimal sampling schemes for multivariate settings remains an important issue for future work.
\newline
For scenarios in which only one specific multivariate method is employed in a single case, the reordering in step 5 has no effect and thus becomes irrelevant. In these situations, our approach reduces to applying the corresponding LDP method and drawing a sample from the obtained multivariate predictive CDF.
\\
Finally, we recall that the established reordering methods for univariate margins as discussed in \ref{ecc} can be mathematically interpreted in terms of empirical copulas \citep{Schefzik2015b}. A similar copula-based framework for the LDP-Reordering method is very likely to hold, while explicit details remain to be worked out.

\subsection{Specific example: Joint postprocessing of wind speed and temperature forecasts at several locations simultaneously using Bivariate EMOS and an ECC-like reordering}\label{ldp.empcop.specific}

\begin{figure*}[p]
\centering
\includegraphics[scale=0.45]{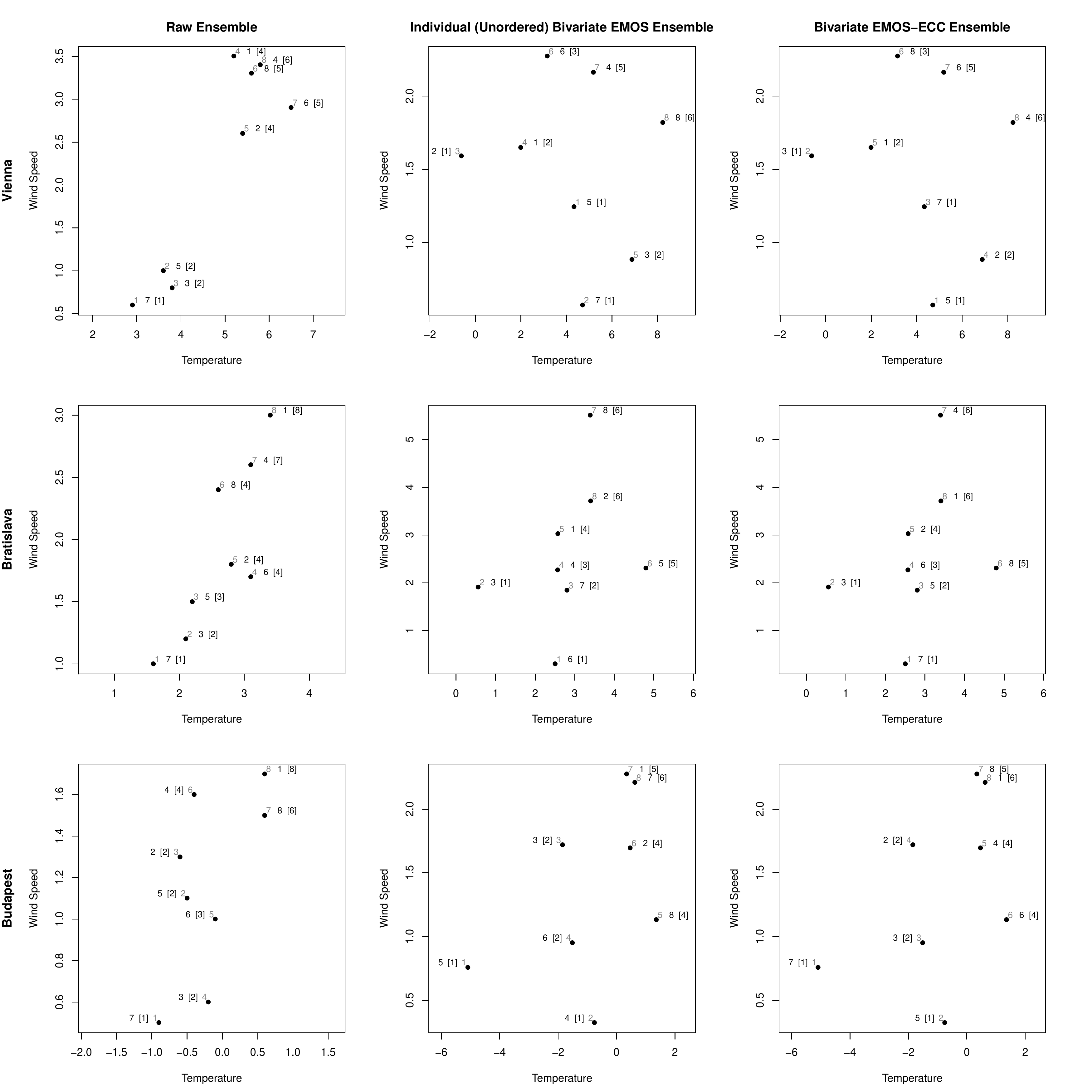}
\caption{Illustration of the Bivariate EMOS-ECC approach based on the example in Subsection \ref{ldp.empcop.specific}, valid 14 December 2013 , 1200 UTC. Each of the $M=8$ ensemble members indicated by a dot is labeled in black with its index and its corresponding bivariate pre-rank according to \eqref{multrank} in brackets. The final rank of each member is given by the gray label. For reasons of clarity, the scales on the axes are different.}
  \label{fig:illus.embed_low}
\end{figure*}
To give an example of an LDP-Reordering approach, we describe the scenario of the real-data case study using the $M=50$-member European Centre for Medium-Range Weather Forecasts (ECMWF) ensemble \citep{Molteni&1996,Buizza2006,ECMWF2012} whose results are presented in Section \ref{case.study}. We consider 24 hour ahead predictions for 10 meter wind speed (W) and surface temperature (T) at Vienna (Austria; Vie24), Bratislava (Slovakia; Br24) and Budapest (Hungary; Bu24), such that $I=2$, $J=3$ and $K=1$, thus facing an $L=I \times J \times K=6$--dimensional setting. The bivariate EMOS approach of \citet{BaranMoeller2015} as described in \ref{lowdim.postpr} is applied to postprocess the wind speed and temperature forecasts jointly at the individual stations of Vienna, Bratislava and Budapest, respectively. Following the partitioning frame of the general LDP-Reordering method, the forecast $\mathbf{x}_m$ of each raw ensemble member $m \in \{1,\ldots,M\}$ is hence made up of
\begin{align*}
&\mathbf{x}_{m}\\&:=(\mathbf{x}_{m}^{\text{Vie24}},\mathbf{x}_{m}^{\text{Br24}},\mathbf{x}_{m}^{\text{Bu24}})\\
&:=(\underbrace{(x_{m}^{T,\,\text{Vie24}},x_{m}^{W,\,\text{Vie24}})}_{\operatorname{case} 1 \operatorname{with} L_1=2},\underbrace{(x_{m}^{T,\,\text{Br24}},x_{m}^{W,\,\text{Br24}})}_{\operatorname{case} 2 \operatorname{with} L_2=2},\underbrace{(x_{m}^{T,\,\text{Bu24}},x_{m}^{W,\,\text{Bu24}})}_{\operatorname{case} 3 \operatorname{with} L_3=2}),
\end{align*}
thus including $C=3$ sub-raw ensemble forecasts or cases, respectively, of dimension $L_c=2$ each, where $c \in \{1,2,3\}$.\\
We later use the whole $M=50$-member ensemble in the evaluation of our method over a longer test period. However, for reasons of clarity and comprehensibility, we demonstrate the procedure for a subset of only $M=8$ raw ensemble members by just taking the first 8 members of the 50-member ECMWF ensemble as the underlying raw ensemble in the following illustration. For the three locations, the left panels of Figure \ref{fig:illus.embed_low} show for each member $m \in \{1,\ldots,8\}$ the respective raw ensemble forecast values $\mathbf{x}_m$ valid 1200 Universal Coordinated Time (UTC) on 14 December 2013, indicated by the dots. Each member is labeled in black with its corresponding index and, in brackets, with its bivariate pre-rank according to \eqref{multrank}, which is exemplarily used as a multivariate characteristic to define a bivariate ranking in the sense of Scheme \ref{scheme.multor} in Subsection \ref{mult.ordering}. The final rank of each raw ensemble member is indicated by the corresponding label in gray. 
\\
The joint postprocessing of the wind speed and temperature ensemble forecasts at each location separately according to the bivariate EMOS method of \citet{BaranMoeller2015} yields bivariate predictive CDFs $F_{\text{Vie24}}$, $F_{\text{Br24}}$ and $F_{\text{Bu24}}$, respectively. Using the rejection sampling algorithm implemented in the R package \texttt{tvmvtnorm} \citep{WilhelmManjunath2010,WilhelmManjunath2014}, we generate a bivariate sample of size $M=8$ from $F_{\text{Vie24}}$, $F_{\text{Br24}}$ and $F_{\text{Bu24}}$, respectively. The mid-panel in Figure \ref{fig:illus.embed_low} shows these samples, which are referred to as the Individual (Unordered) Bivariate EMOS ensemble, again labeled in black with their corresponding indices and bivariate pre-ranks (in brackets) according to \eqref{multrank}, which are derived analogously as for the raw ensemble, and labeled in gray with their final ranks.
\\
The desired reordered postprocessed ensemble is finally based on an ECC-like notion using the raw ensemble as a dependence template, that is, we have $\mathbf{z}=\mathbf{x}$ and $N=M$ in the general LDP-Reordering approach introduced in Subsection \ref{ldp.empcop}. This ensemble is referred to as the Bivariate EMOS-ECC ensemble and is shown in the right panel of Figure \ref{fig:illus.embed_low}. Again, each ensemble member is labeled in black with its index and its bivariate pre-rank (in brackets), and in gray with its final rank.
\\
The pairs of ensemble member forecasts for wind speed and temperature forecasts and their corresponding bivariate pre-ranks (labeled in black in brackets), and thus also their respective final ranks (labeled in gray), for Vienna, Bratislava and Budapest, respectively, are exactly the same for both the Individual (Unordered) Bivariate EMOS (middle) and the Bivariate EMOS-ECC (right) ensemble. However, the members of the Bivariate EMOS-ECC approach have other indices than those of the Individual (Unordered) Bivariate EMOS ensemble, following the bivariate pre-rank structure in the raw ensemble. A final Bivariate EMOS-ECC ensemble member $m$ for Vienna, Bratislava and Budapest, respectively, should have a bivariate pre-rank that is near to or comparable to the bivariate pre-rank which the raw ensemble member with the index $m$ has. The same should consequently hold for the final rank then. An exact confirmity of the bivariate pre-rank structure and the final rank pattern, respectively, of the raw and the Bivariate EMOS-ECC ensemble is not always possible, as in the samples drawn from the Bivariate EMOS distributions, other bivariate pre-ranks with different frequencies than in the raw ensemble might occur. However, the Bivariate EMOS-ECC approach tries to conserve the given structure to a great extent.
\\
For instance, in our example for Budapest in the third row of Figure \ref{fig:illus.embed_low}, the raw ensemble member with the index 7 has the lowest bivariate pre-rank (left panel), namely 1, and thus the final rank 1. This is respected by the Bivariate EMOS-ECC ensemble (right panel), whose member with the index 7 also has a bivariate pre-rank of 1 and hence also a final rank of 1. In contrast, this property gets drastically lost for the Individual (Unordered) Bivariate EMOS ensemble (mid-panel), for which the member with the index 7 has a bivariate pre-rank of 6, which is the highest in this case, and thus also the highest final rank, namely 8. So both the bivariate pre-rank and the final rank of the Individual (Unordered) Bivariate EMOS ensemble member with index 7 are unsuitably high with respect to the raw ensemble.

\section{Case study}\label{case.study}

\subsection{Setting and implementation details}\label{data.set}

We now evaluate the Bivariate EMOS-ECC approach in a case study, following the setting in the example in Subsection \ref{ldp.empcop.specific}. That is, we consider 24 hour ahead joint 10 meter wind speed and surface temperature forecasts, all valid 1200 UTC, at Vienna (Austria), Bratislava (Slovakia) and Budapest (Hungary) simultaneously, provided by the $M=50$-member European Centre for Medium-Range Weather Forecasts (ECMWF) core ensemble \citep{Molteni&1996,Buizza2006,ECMWF2012}. Our test period ranges from 1 January 2003 to 31 December 2013, comprising in total 3983 instances after the removal of dates with missing forecast and/or observation data. The temperature observations show very high spatial correlations among the three locations, while the spatial dependencies of the wind speed observations are highly to moderately pronounced. Inter-variable correlations between the temperature and wind speed observations are existent, but are not that marked compared to the spatial dependencies. All observational correlation structures are mirrored reasonably well in the raw ensemble. 
\\
We employ both univariate and bivariate EMOS methods to postprocess the wind speed and temperature predictions. In both cases, the respective model parameters are estimated based on training data only from the individual station of interest, thus yielding a distinct set of parameters for each location. We use a sliding window of 50 days as a training period for parameter estimation in both implementations.
\\
Univariate postprocessing, which applies to each weather variable separately, is performed via the EMOS approaches \eqref{emos.gaussian} for temperature and \eqref{emos.truncnormal} for wind speed, respectively, as discussed in \ref{emos}. Optimization in the univariate EMOS approaches is performed with respect to the continuous ranked probability score (CRPS), because if a closed-form expression of the CRPS is available, which is the case for the models \eqref{emos.gaussian} and \eqref{emos.truncnormal}, this usually yields better results than optimizing with respect to the logarithmic score. Concerning the joint postprocessing of wind speed and temperature forecasts, we use the bivariate EMOS approach of \citet{BaranMoeller2015} as presented in \ref{lowdim.postpr}.  For this method, we follow \citet{BaranMoeller2015} and estimate the model parameters by optimizing the mean logarithmic score, as there is no closed-form expression available for the CRPS in the case of a bivariate truncated normal distribution. 
\\
As we aim to compare the different approaches in form of ensemble forecasts, we need to discretize the (either univariate or bivariate) postprocessed predictive distributions. In the univariate case, we employ the EMOS-R ensemble, which is generated by using quantization scheme (R) from \ref{emos}, that is, we just draw a random sample from each marginal predictive distribution. Similarly, the samples from the Bivariate EMOS predictive distributions are also drawn in a random manner by using the rejection sampling method implemented in the R package {\texttt{tvmvtnorm}} \citep{WilhelmManjunath2010,WilhelmManjunath2014}. Note that for univariate settings, samples according to scheme (Q) in \ref{emos} are expected to yield better results than samples based on scheme (R), as equidistant quantiles can be considered kind of an optimal choice for univariate sampling \citep{Broecker2012,GrafLuschgy2000}. In contrast, there appears to be no obvious optimal sampling strategy for multivariate settings, such that for a fairer comparison, we only consider random sampling both in the univariate and the bivariate case here.
\\
Having performed the discretization, the (uni- or bivariate) samples get reordered according to the dependence structure in the raw ensemble. In the univariate case, this is done by applying the ECC approach by \citet{Schefzik&2013} as discussed in \ref{ecc}, finally leading to the EMOS-ECC-R ensemble. For the Bivariate EMOS samples, we perform the reordering via the ECC-like procedure illustrated in Subsection \ref{ldp.empcop.specific}, where we study variants based on the different bivariate ranking characteristics according to the bivariate pre-rank (BivPR) in \eqref{multrank}, the average pre-rank (AvPR) in \eqref{avgrank} and the signed Euclidean norm (SEN) in \eqref{norms.signbased} as described in Subsection \ref{mult.ordering}. We refer to the corresponding ensembles as the Bivariate EMOS-ECC ensemble (BivPR), the  Bivariate EMOS-ECC ensemble (AvPR) and the Bivariate EMOS-ECC ensemble (SEN), respectively, in what follows.   
\\
In a nutshell, we compare the predictive performances of the following reference ensembles:
\begin{itemize}
\item{the unprocessed raw ensemble,}
\item{the EMOS-ECC-R ensemble,}
\item{the Individual (Unordered) Bivariate EMOS ensemble consisting of the unordered bivariate samples obtained from the Bivariate EMOS postprocessing, by analogy with the situation illustrated in the mid-panel of Figure \ref{fig:illus.embed_low}, and}
\item{three different Bivariate EMOS-ECC ensembles, namely
\begin{itemize}
\item{the Bivariate EMOS-ECC ensemble (BivPR), }
\item{the Bivariate EMOS-ECC ensemble (AvPR), and}
\item{the Bivariate EMOS-ECC ensemble (SEN).}
\end{itemize}
}
\end{itemize}

\subsection{Evaluation tools}\label{app.verif}

Following \citet{Gneiting&2007}, a predictive distribution or an ensemble forecast, respectively, should be as sharp as possible, but subject to being calibrated, where calibration refers to the statistical consistency between forecasts and observation. There are several verification tools to evaluate the predictive performances of our reference ensembles \citep{Wilks2011}. To describe the evaluation tools used in this paper, let
\begin{equation*}
\mathbf{x}_1:=(x_1^1,\ldots,x_1^L),\ldots,\mathbf{x}_N:=(x_N^1,\ldots,x_N^L) \in \mathbb{R}^L
\end{equation*}
denote an $N$-member ensemble forecast and
\begin{equation*}
\mathbf{y}:=(y_1,\ldots,y_L) \in \mathbb{R}^L
\end{equation*}
an observation in the case of an individual forecast instance.\\
We assess calibration by computing the rank of the observation $\mathbf{y}$ within the pooled set $\{\mathbf{x}_1,\ldots,\mathbf{x}_N,\mathbf{y}\}$ for all individual forecast cases and plotting the respective rank histogram. In the case of a calibrated ensemble forecast, the ranks are uniformly distributed on $\{1,\ldots,N+1\}$, thus yielding a flat rank histogram. In univariate settings, the common ranking can be employed, leading to the verification rank histogram \citep{Anderson1996,Talagrand&1997,Hamill2001}. For multivariate quantities, which are our focus, we use the multivariate, band depth and average ranking concepts and histograms, respectively \citep{Gneiting&2008,Thorarinsdottir&2013}. A quantitative measure for the deviation from a uniform distribution on $\{1,\ldots,N+1\}$ is the reliability index $\Delta$ given by
\begin{equation*}
\Delta:=\sum\limits_{r=1}^{N+1} \left| \rho_r - \frac{1}{N+1} \right|,
\end{equation*}
where $\rho_r$ denotes the relative frequency of rank $r$ \citep{DelleMonache&2006}. Depending on if we use the multivariate, band depth or average rank concept for the ranking, we refer to $\Delta_{\operatorname{MR}}$, $\Delta_{\operatorname{BDR}}$ and $\Delta_{\operatorname{AvR}}$, respectively, for the corresponding histograms in what follows.
\\
The overall forecast skill is evaluated via proper scoring rules \citep{GneitingRaftery2007}, which give a single number assessing calibration and sharpness simultaneously. We take scoring rules to be negatively oriented here, meaning that the lower the score the better the predictive performance. An established proper scoring rule for univariate quantities is the continuous ranked probability score (CRPS) \citep{GneitingRaftery2007,Hersbach2000}. Its analog for multivariate quantities is the energy score (ES) \citep{Gneiting&2008}, which in case of an ensemble forecast is computed via
\begin{align*}
\operatorname{ES}(\mathbf{x}_1,\ldots,\mathbf{x}_N;\mathbf{y})=\frac{1}{N}\sum\limits_{n=1}^{N}||\mathbf{x}_n-\mathbf{y}||-\frac{1}{2 N^2}\sum\limits_{\nu=1}^{N}\sum\limits_{n=1}^{N}||\mathbf{x}_{\nu}-\mathbf{x}_n||,
\end{align*} 
where $||\cdot||$ denotes the Euclidean norm. The ES might fail to detect misspecifications in the correlation structure \citep{PinsonTastu2013,ScheuererHamill2014}. In contrast, the variogram score (VS) \citep{ScheuererHamill2014} is more sensitive in this respect. Here, we use the VS in the form
\begin{align*}
\operatorname{VS}(\mathbf{x}_1,\ldots,\mathbf{x}_N;\mathbf{y})=\sum\limits_{\ell=1}^{L}\sum\limits_{\lambda=1}^{L} \left(\big \vert y_{\ell}-y_{\lambda}\big \vert^{\frac{1}{2}}-\frac{1}{N}\sum\limits_{n=1}^{N}\big\vert x_{n}^{\ell}-x_{n}^{\lambda} \big \vert ^{\frac{1}{2}}\right)^2,
\end{align*}
being the special case of a $0.5$-VS for ensemble forecasts in the general formulation in \citet{ScheuererHamill2014}, which additionally includes optional non-negative weights. However, as inter-variable dependencies, for which a weighting is not obvious or makes sense \citep{ScheuererHamill2014}, are part of our considerations, we omit the weights here.\\
In our case study, we report average scores over all forecast cases within a specific test period. As we deal with weather variables having distinct units and magnitudes (wind speed and temperature), the forecasts and observations are standardized for an evaluation via scores here. Precisely, a quantity $z$ is transformed to a standardized quantity $z^\ast$ via $z^\ast:=(z-\mu)/\sigma$, where $\mu$ is taken to be the climatological mean over all available observations in the data set for the respective quantity at the specific location, and $\sigma$ the corresponding climatological standard deviation.

\subsection{Results}\label{results}

\begin{table*}[t]
\caption{Average energy scores (ES) and variogram scores (VS) for 24 hour ahead temperature and wind speed forecasts at Vienna, Bratislava and Budapest jointly over 3983 test days during the period from 1 January 2003 to 31 December 2013. The scores for all ensembles except for the raw ensemble are averaged over 200 different samples.}\label{tab:scores}
\begin{center}
\begin{tabular}{lcc}
\hline \hline
&ES&VS\T\B\\
\hline
ECMWF Raw Ensemble&1.014&3.348\vspace{0.15 cm}\\
EMOS-ECC-R Ensemble&0.773&2.467\vspace{0.15 cm}\\
Individual (Unordered) Bivariate EMOS Ensemble&0.776&2.471\vspace{0.15 cm}\\
Bivariate EMOS-ECC Ensemble (BivPR)&0.775&2.440\\
Bivariate EMOS-ECC Ensemble (AvPR)&0.775&2.439\\
Bivariate EMOS-ECC Ensemble (SEN)&0.775&2.437\\
\hline \hline
\end{tabular}
\end{center}
\end{table*}

\begin{table*}[t]
\caption{Reliability indices $\Delta_{\operatorname{MR}}$, $\Delta_{\operatorname{BDR}}$ and $\Delta_{\operatorname{AvR}}$ corresponding to the multivariate, band depth and average rank histograms, respectively, in Figure \ref{fig:rankhist}, evaluating 24 hour ahead temperature and wind speed forecasts at Vienna, Bratislava and Budapest jointly over 3983 test days during the period from 1 January 2003 to 31 December 2013. The results for all ensembles except for the raw ensemble are based on aggregations over 200 different samples.}\label{tab:discr.ind}
\begin{center}
\begin{tabular}{lccc}
\hline \hline
&$\Delta_{\operatorname{MR}}$&$\Delta_{\operatorname{BDR}}$&$\Delta_{\operatorname{AvR}}$\T\B\\
\hline
ECMWF Raw Ensemble&0.622&1.261&0.397\vspace{0.15 cm}\\
EMOS-ECC-R Ensemble&0.181&0.158&0.052\vspace{0.15 cm}\\
Individual (Unordered) Bivariate EMOS Ensemble&0.108&0.191&0.306\vspace{0.15 cm}\\
Bivariate EMOS-ECC Ensemble (BivPR)&0.150&0.121&0.071\\
Bivariate EMOS-ECC Ensemble (AvPR)&0.138&0.135&0.093\\
Bivariate EMOS-ECC Ensemble (SEN)&0.111&0.124&0.076\\
\hline \hline
\end{tabular}
\end{center}
\end{table*}

\begin{figure*}[p]
\centering
\subfigure[Multivariate rank histograms]{\includegraphics[scale=0.4]{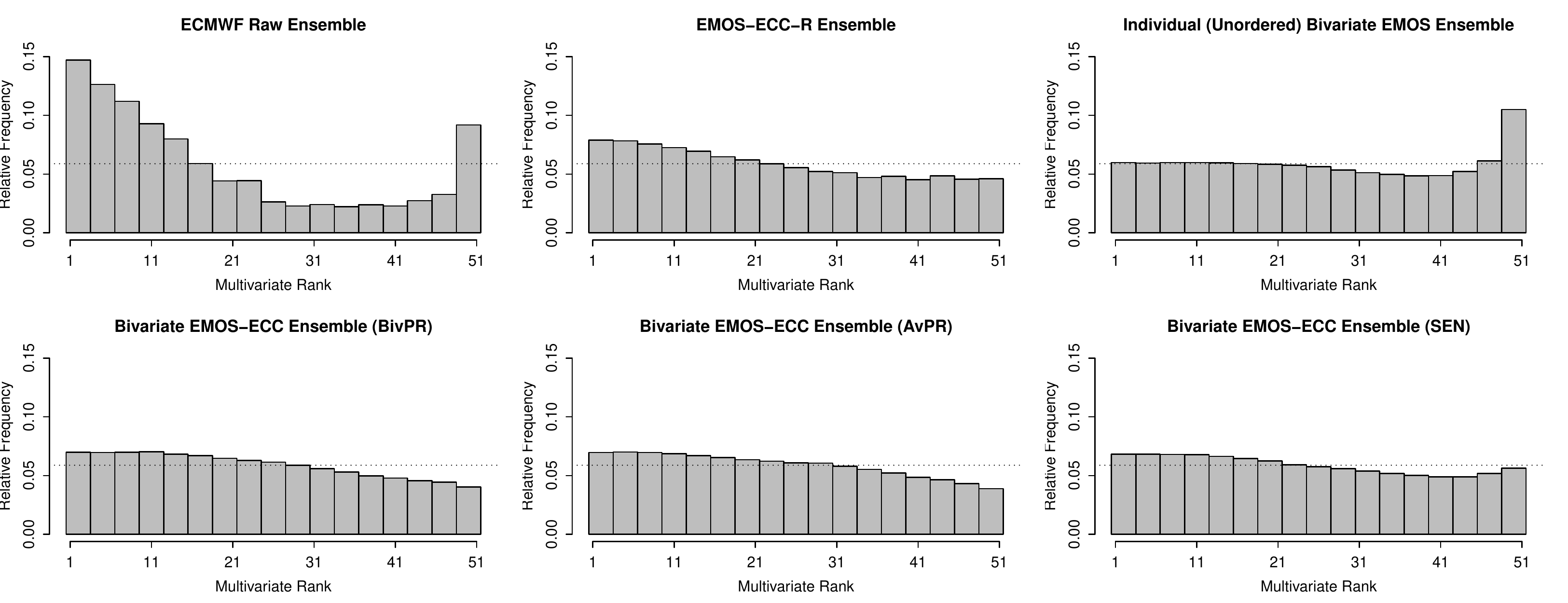}}
\vspace{-1mm} \\
\subfigure[Band depth rank histograms]{\includegraphics[scale=0.4]{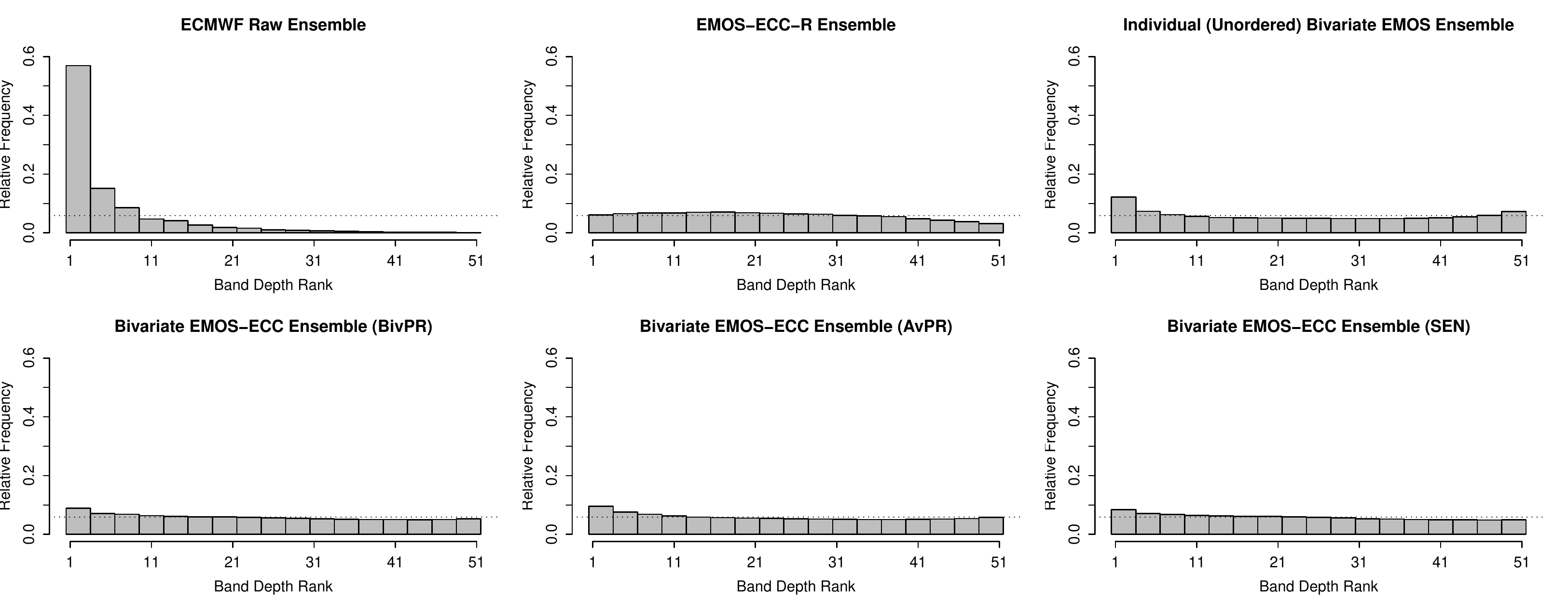}}
\vspace{-1mm} \\
\subfigure[Average rank histograms]{\includegraphics[scale=0.4]{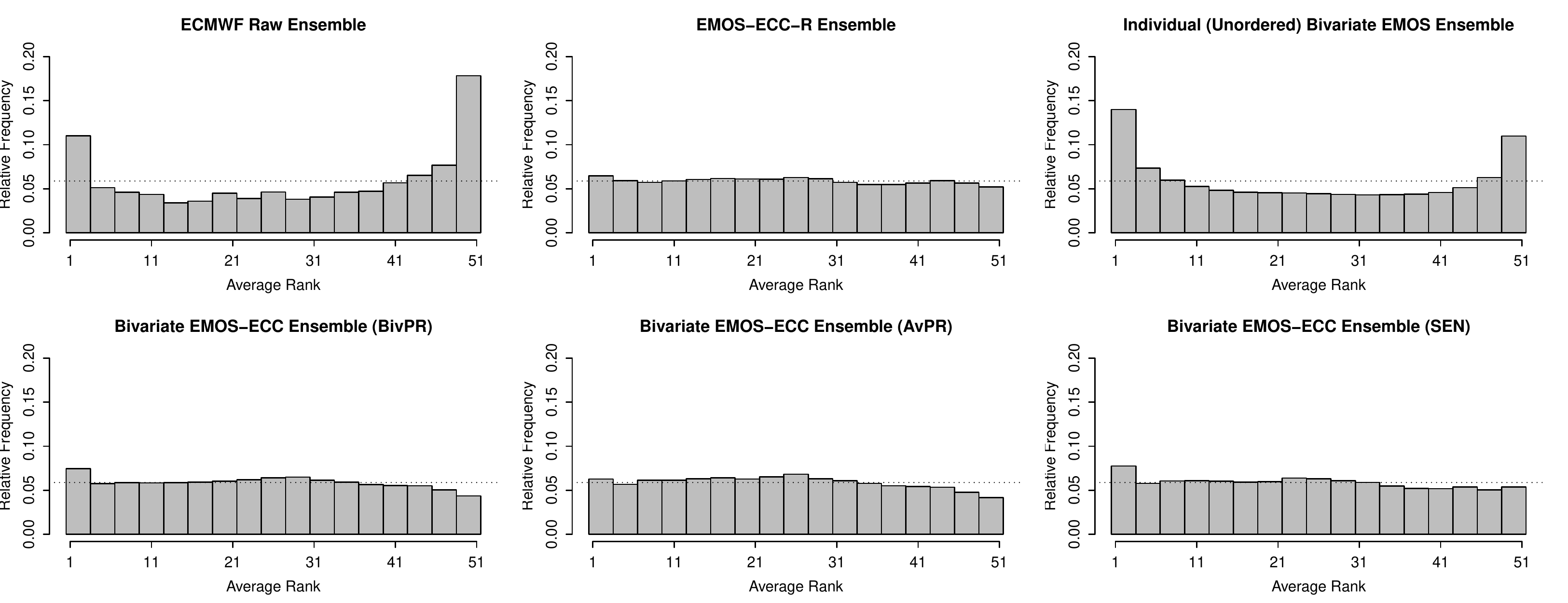}}
\caption{(a) Multivariate, (b) band depth and (c) average rank histograms for 24 hour ahead temperature and wind speed forecasts at Vienna, Bratislava and Budapest jointly over 3983 test days during the period from 1 January 2003 to 31 December 2013. The results for all ensembles except for the raw ensemble are aggregated over 200 different samples.}
  \label{fig:rankhist}
\end{figure*}
The average energy scores (ES) and variogram scores (VS), respectively, as overall performance measures for the different ensembles are shown in Table \ref{tab:scores}. Calibration is checked via the multivariate, band depth and average rank histograms, respectively, in Figure \ref{fig:rankhist}, with the corresponding reliability indices in Table \ref{tab:discr.ind}. To account for the random component in the sampling procedures, the scores for the EMOS-ECC-R, the Individual (Unordered) Bivariate EMOS and the three Bivariate EMOS-ECC ensembles are aggregated and reported as averages, respectively, over 200 different samples. As our focus is on the multivariate evaluation, we do not explicitly show results for the univariate EMOS postprocessing, but mention briefly that EMOS clearly outperforms the raw ensemble, both for temperature and wind speed.
\\
With respect to the multivariate evaluation, the raw ensemble exhibits a U-shaped multivariate and average rank histogram and a left-skewed band depth rank histogram, each with a strong deviation from uniformity. It is thus clearly underdispersive and consequently uncalibrated, hence calling for postprocessing. Indeed, all postprocessed ensembles outperform the raw ensemble in terms of scores and reliability indices.
\\
The four postprocessed ensembles respecting the dependence pattern of the raw ensemble (that is, the EMOS-ECC-R and the three Bivariate EMOS-ECC ensembles) perform well. In contrast, the postprocessed Individual (Unordered) Bivariate EMOS ensemble not accounting for dependencies reveals weaknesses, although it also outperforms the raw ensemble. Considering the ES only, these shortcomings do not become that visible, as the differences between the ES values of the postprocessed ensembles are rather minor. While the ES is suitable to emphasize the benefits of postprocessing itself, it fails to detect misspecifications in the correlation structures \citep{PinsonTastu2013,ScheuererHamill2014}. The VS has shown to be more sensitive in this regard \citep{ScheuererHamill2014}. Indeed, in terms of the VS, all postprocessed ensembles respecting dependencies perform better than the Individual (Unordered) Bivariate EMOS ensemble. The unsatisfactory performance of the Individual (Unordered) Bivariate EMOS ensemble is further witnessed by the U-shape of the corresponding band depth and average rank histogram, respectively, which indicates an underestimation of the correlation structure \citep{Thorarinsdottir&2013}, also resulting in poor reliability indices corresponding to these histograms. Note that with respect to the multivariate rank histogram, the Individual (Unordered) Bivariate EMOS ensemble admittedly performs even best, showing the lowest reliability index, if only slightly. However, similar to the ES, the multivariate rank histogram does not appear to be suitable to detect inadequate correlation structures, especially in higher dimensions \citep{Thorarinsdottir&2013}. In a nutshell, the ECC-like reordering proves to be indeed essential in our new approach.
\\
Concerning the three Bivariate EMOS-ECC ensembles in comparison among themselves, each ensemble performs well in principle, and there are no differences in terms of the ES. With respect to the VS, the three Bivariate EMOS-ECC ensemble perform well on nearly the same level, with the Bivariate EMOS-ECC (SEN) ensemble being slightly  best. These findings are also reflected in the rank histograms and the discrepancy indices. As far as band depth and average rank histograms are concerned, the three Bivariate EMOS-ECC ensembles perform more or less equally well, while in the case of the multivariate rank histogram, the Bivariate EMOS-ECC (SEN) ensemble performs better than the other two Bivariate EMOS-ECC ensembles in a more pronounced way. This appears to be plausible, as the ranking according to the signed Euclidean norm has been specifically tailored to the situation in this case study, whereas the rankings follwing the multivariate or average pre-rank notions are more general, but as shown nevertheless appropriate.  
\\
With respect to the ES, the EMOS-ECC-R ensemble relying on univariate EMOS postprocessing very slightly outperforms the three Bivariate EMOS-ECC ensembles based on bivariate EMOS postprocessing. In contrast, the Bivariate EMOS-ECC ensembles outperform EMOS-ECC-R in a more pronounced way in terms of the VS being more sensitive to correlations. Furthermore, the Bivariate EMOS-ECC ensembles exhibit a bit more skill than the EMOS-ECC-R ensemble considering the multivariate and band depth rank histograms and the corresponding reliability indices. In our case study here, the benefits of using bivariate EMOS instead of univariate EMOS postprocessing are apparent, but may not come into effect that much, as the correlations between wind speed and temperature are existent, but not severely pronounced at each of the single locations. In contrast, the spatial correlations between Vienna, Bratislava and Budapest are much stronger for both weather variables, such that the ECC-like reordering plays a more important role here than the bivariate modeling. Moreover, recall that the bivariate EMOS postprocessing has been performed by optimizing with respect to the logarithmic score, while the univariate EMOS methods have employed an optimization in terms of the CRPS, which usually yields better scores. If there was an efficient way to optimize with respect to the CRPS in the case of a bivariate truncated normal distribution, for which no closed-form expression for the CRPS is available, the results for the bivariate EMOS-based approaches could likely be improved further. In summary, the bivariate EMOS modeling makes sense and is useful here, but the need therefore would be again more important if the inter-variable correlations were more pronounced. Note also that the bivariate EMOS method does not require considerably more computational effort than the univariate EMOS postprocessing for our purposes here.

\section{Discussion}\label{discussion}

We have shown how to combine parametric multi-(low-)dimensional ensemble postprocessing approaches with non-parametric reordering notions based on dependence templates, leading to the novel LDP-Reordering method.
\\
In this context, the challenge of defining a multivariate ranking arises, and we have considered three options to address this issue: the multivariate pre-rank (MultPR), the average pre-rank (AvPR) and the signed Euclidean norm (SEN). The MultPR- and AvPR-based rankings are rather general and may often be subject to a random component due to the random allocation of possible ties. In contrast, the SEN-based ranking concept has been specifically designed for the frame of wind speed and temperature forecasts as examined in our case study and is likely to involve fewer ties than the other two concepts and thus confronted with less randomization. More sophisticated ranking concepts for multivariate settings, maybe tailored to specific settings one is interested in, and/or more sophisticated allocation methods in case of ties could be developed in the future.
\\
Moreover, further research about how to sample from a multivariate distribution is strongly encouraged. In the univariate case, equidistant quantiles provide a canonical and to some extent also optimal choice when drawing a sample \citep{Broecker2012,GrafLuschgy2000}. However, similar optimality results are not available for the multivariate situation. As a first attempt into this direction, one could for instance seek a generalization of the concept of equidistance for multidimensional settings, which is however not obvious at first sight, and check for optimality.
\\
From a mathematical point of view, it remains to explicitly relate the LDP-Reordering approach to the (empirical) copula framework, similarly as shown for the univariate reordering approaches \citep{Schefzik2015b}. In the LDP-Reordering method, we seek a multivariate predictive CDF with fixed multivariate margins. If all marginals CDFs $F_1,\ldots,F_L$ are univariate, the celebrated Sklar's theorem \citep{Sklar1959,Nelsen2006} allows for a construction of a multivariate CDF $H$ using a suitable copula function $C$, which models the dependence, via the decomposition
\begin{equation*}
H(y_1,\ldots,y_L)=C(F_1(y_1),\ldots,F_L(y_L))
\end{equation*}
for $y_1,\ldots,y_L \in \mathbb{R}$.  It would be nice if a result similar to Sklar's theorem could be stated in the case of multivariate (not necessarily univariate) margins, but this appears to be a non-trivial open question for further research. For some special cases, such as for fixed bi- or trivariate margins, theory is available to some extent \citep{Joe1997,Joe2014}.
\\
From an applied perspective, we have employed the bivariate EMOS approach of \citet{BaranMoeller2015} to postprocess wind speed and temperature ensemble forecasts jointly, but at individual stations. The spatial correlations have been addressed by a reordering in an ECC-like manner using the raw ensemble forecasts as a dependence template, based on the three different multivariate ranking concepts mentioned above. All Bivariate EMOS-ECC ensembles have shown a good predictive performance. The benefits of using both a multivariate reordering and a bi- or multivariate parametric postprocessing to obtain the marginal distributions have become clear in our case study. However, the latter step may be even more powerful in other settings in which the inter-variable correlation structures are more pronounced.  For instance, one could work with the bivariate EMOS approach of \citet{Schuhen&2012} to postprocess $(u,v)$-wind forecast vectors or with the Gaussian copula method of \citet{Moeller&2013} to postprocess temperature and pressure forecasts jointly, as for both pairs of variables, the dependence patterns might be expected to be stronger. The design of further case studies to examine this on the basis of a suitable data set is highly desirable. Another further step is to perform case studies using an observation-based multivariate reordering according to the Schaake shuffle \citep{Clark&2004} or modifications thereof \citep{Schefzik2015c}, instead of employing the ECC-like reordering based on a template of raw forecasts.

%%%%%%%%%%%%%%%%%%%%%%%%%%%%%%%%%%%%%%%%%%%%%%%%%%%%%%%%%%%%%%%%%%%%%
% ACKNOWLEDGMENTS
%%%%%%%%%%%%%%%%%%%%%%%%%%%%%%%%%%%%%%%%%%%%%%%%%%%%%%%%%%%%%%%%%%%%%
%
\section*{Acknowledgments}
I gratefully acknowledge the support of the Klaus Tschira Foundation and the support of the VolkswagenStiftung under the project 
``Mesoscale Weather Extremes: Theory, Spatial Modeling and Prediction''. Initial work on this paper was done during my time as Ph.D. student at Heidelberg University, funded by Deutsche Forschungsgemeinschaft through Research Training Group (RTG) 1953. I thank Tilmann Gneiting for valuable comments and suggestions. The forecast data used in the case study have been made available by the European Centre for Medium-Range Weather Forecasts. I thank Stephan Hemri for help with the data, and S\'{a}ndor Baran and Michael Scheuerer for R code.

%%%%%%%%%%%%%%%%%%%%%%%%%%%%%%%%%%%%%%%%%%%%%%%%%%%%%%%%%%%%%%%%%%%%%
% APPENDIXES
%%%%%%%%%%%%%%%%%%%%%%%%%%%%%%%%%%%%%%%%%%%%%%%%%%%%%%%%%%%%%%%%%%%%%
%
% Use \appendix if there is only one appendix.
%\appendix

% Use \appendix[A], \appendix}[B], if you have multiple appendixes.
%\appendix[A]

%% Appendix title is necessary! For appendix title:
%\appendixtitle{}

%%% Appendix section numbering (note, skip \section and begin with \subsection)
% \subsection{First primary heading}

% \subsubsection{First secondary heading}

% \paragraph{First tertiary heading}

%% Important!
%\appendcaption{<appendix letter and number>}{<caption>} 
%must be used for figures and tables in appendixes, e.g.,
%
%\begin{figure}
%\noindent\includegraphics[width=19pc,angle=0]{figure01.pdf}\\
%\appendcaption{A1}{Caption here.}
%\end{figure}
%
% All appendix figures/tables should be placed in order AFTER the main figures/tables, i.e., tables, appendix tables, figures, appendix figures.
%
%%%%%%%%%%%%%%%%%%%%%%%%%%%%%%%%%%%%%%%%%%%%%%%%%%%%%%%%%%%%%%%%%%%%%
% REFERENCES
%%%%%%%%%%%%%%%%%%%%%%%%%%%%%%%%%%%%%%%%%%%%%%%%%%%%%%%%%%%%%%%%%%%%%
% Make your BibTeX bibliography by using these commands:
%\bibliographystyle{ametsoc2014}

\bibliographystyle{plainnat}
\bibliography{Biblio_LDP_Reordering_RomanSchefzik}

\end{document}